\documentclass[pre,10pt,twocolumn,superscriptaddress,floatfix]{revtex4-2}
\usepackage[utf8]{inputenc}
\usepackage[T1]{fontenc}
\usepackage{hyperref}
\usepackage{amsmath,amsthm,amssymb} 
\usepackage{graphicx}

\begin{document}
\renewcommand{\acknowledgmentsname}{Acknowledgements}

\title{Phase and amplitude synchronisation in power-grid frequency fluctuations in the Nordic Grid}

\author{Leonardo~Rydin~Gorj\~ao}
\email{l.rydin.gorjao@fz-juelich.de}
\affiliation{Forschungszentrum J\"ulich, Institute for Energy and Climate Research - Systems Analysis and Technology Evaluation (IEK-STE), 52428 J\"ulich, Germany}
\affiliation{Institute for Theoretical Physics, University of Cologne, 50937 K\"oln, Germany}

\author{Luigi~Vanfretti}
\affiliation{Electrical, Computer, and Systems Engineering, Rensselaer Polytechnic Institute, 8024 Troy, New York, United States}

\author{Dirk~Witthaut}
\affiliation{Forschungszentrum J\"ulich, Institute for Energy and Climate Research - Systems Analysis and Technology Evaluation (IEK-STE), 52428 J\"ulich, Germany}
\affiliation{Institute for Theoretical Physics, University of Cologne, 50937 K\"oln, Germany}

\author{Christian~Beck}
\affiliation{School of Mathematical Sciences, Queen Mary University of London, United Kingdom}

\author{Benjamin~Sch\"afer}
\email{b.schaefer@qmul.ac.uk}
\affiliation{School of Mathematical Sciences, Queen Mary University of London, United Kingdom}

\begin{abstract}
Monitoring and modelling the power grid frequency is key to ensuring stability in the electrical power system.
Many tools exist to investigate the detailed deterministic dynamics and especially the bulk behaviour of the frequency. 
However, far less attention has been paid to its stochastic properties, and there is a need for a cohesive framework that couples both short-time scale fluctuations and bulk behaviour.
Moreover, commonly assumed uncorrelated stochastic noise is predominantly employed in modelling in energy systems.
In this publication, we examine the stochastic properties of six synchronous power-grid frequency recording with high-temporal resolution of the Nordic Grid from September 2013, focusing on the increments of the frequency recordings.
We show that these increments follow non-Gaussian statistics and display spatial and temporal correlations.
Furthermore, we report two different physical synchronisation phenomena: a very short timescale phase synchronisation ($<2\,$s) followed by a slightly larger timescale amplitude synchronisation ($2\,$s--$5\,$s).
Overall, these results provide guidance on how to model fluctuations in power systems.
\end{abstract}

\maketitle

\begin{figure*}[t]
    \includegraphics[width=\linewidth]{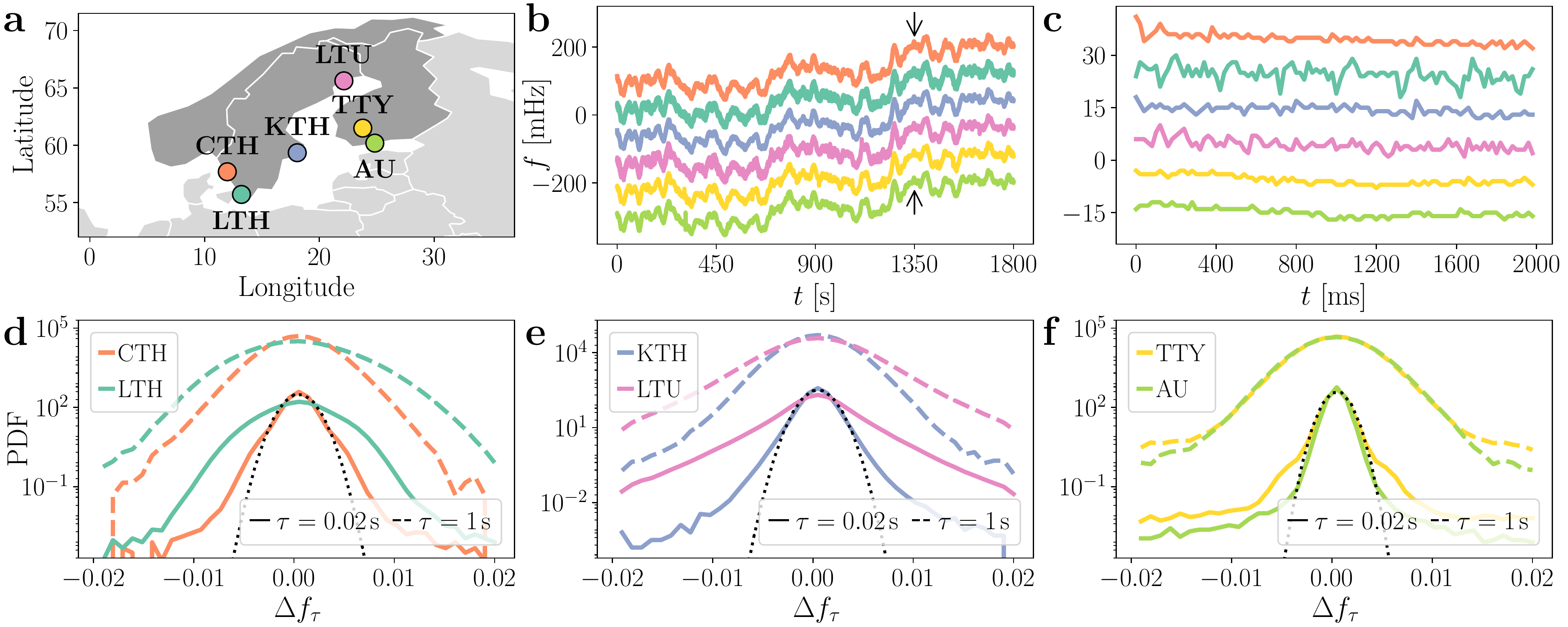}
    \caption{Overview of the data. The six power-grid frequency recordings in the Nordic Grid from 2013 show very distinct increment statistics.
    \textbf{a} Approximate locations of the recordings across the Nordic Grid: CTH, LTH, KTH, LTU, TTY, and AU. 
    \textbf{b} Excerpts of the recordings in a $30$ minutes time scale, displaced vertically.
    \textbf{c} Zoom into the arrows of panel \textbf{b} of a total length of $2\,$seconds, displaced vertically.
    \textbf{d}-\textbf{f} display the probability density functions (PDF) of the increments $\Delta f_\tau$ at the shortest increment lag $\tau=0.02\,$s and at $\tau=1\,$s, in a vertical logarithmic scale, alongside a normal distribution (which is an inverted parabola in a vertical logarithmic scale) with equal variance for the first recording at $\tau=0.02\,$s, for comparison.
    PDFs are vertically displaced for clarity. 
    All recordings are synchronous and have a sampling time of $0.02~\!$s.}
    \label{fig:trajectories}
\end{figure*}

\section{Introduction}

The power-grid frequency is a key indicator of the stability of electric power systems. 
It weighs in the balance of power generation and consumption as well as the operation of each component of the grid and thus serves as the main signal used for balancing~\cite{Machowski2020}.
In theory, during steady operation, the frequency is the same throughout the grid and all generators have a fixed phase difference that essentially determines the real power flows. 
Yet, perfect phase locking is only an approximation to the operation of real power grids subject to numerous external perturbations. 
For instance, perturbations, such as the tripping of important lines or the loss of a major generating unit, can cause inter-area oscillations corresponding to the normal modes of the grid around a quasi-equilibrium~\cite{Klein1991}.
Large perturbations can even lead to the complete loss of synchrony between different parts of the power grid, eventually leading to blackouts~\cite{ucte2007final}.
While luckily these large contingencies occur seldom, small fluctuations---ambient noise---are always present. These, nuisance fluctuations, are ubiquitous, and can help us understand a wide range of characteristics of power-grid frequency, such as the role of inertia or renewable energy intermittency~\cite{Anvari16,Weber2019}.
This article focuses on analysing, characterising, and explaining deviations caused by these ambient perturbations and the ability of the grid to relax to synchrony afterwards.

Frequency dynamics and synchronisation are essential aspects of power system operation and thus intensively studied in the literature. 
The ongoing energy transition strengthened the interest in these topics, as synchronous generators are replaced by inverter-based power sources~\cite{Ulbig2014,Milano2018}.
For instance, recent years saw enormous progress in the mathematical theory of synchronisation~\cite{Dorfler2014}, leading to the derivation of a variety of rigorous stability conditions~\cite{Dorfler2012,Dorfler2013}.
An important practical research topic is the dynamics and design of inverter-based power grids lacking the inertia provided by large synchronous machines~\cite{Schiffer2014,Colombino2019,Homan2021,Farmer2021}.
Another key area is the development of detailed simulation models to study frequency dynamics and synchronisation for actual grid layouts and contingency situations~\cite{Chiang1987,Faruque2015}. 
Power hardware-in-the-loop then allows to investigate and test actual equipment coupled with real-time simulations~\cite{Lu2007,Lauss2015}. 
A common scenario in such simulation-based studies is the dynamics after a sudden large perturbation, modelling a fault (see, e.g. Ref.~\cite{Ulbig2014}).
Does the grid remain stable? And how long does it take to relax to a steady synchronous operation? 
The current manuscript adopts a very different approach to this synchronisation in power-grid, focusing on the analysis of synchronous frequency measurements in the presence of ambient noise. 
We will extract the essential scales of synchronisation in space and time from frequency time series in a model-free way.

Non-linear time series analysis has been applied to investigate various aspects of power-grid systems, including using some stochastic analysis methods~\cite{Messina2009}, particularly the application of the Hilbert--Huang transform (empirical mode decomposition)~\cite{Feldman2011}, wavelet-based analysis~\cite{Rueda2011}, or simply investigating the spectral density~\cite{Vanfretti2013}. 
The spatio-temporal dynamics of the power system have been studied in more detail, e.g. by revealing inter- and intra-area oscillations and eigenfrequencies. 
In the Nordic Grid, in particular, eigenfrequencies~\cite{Uhlen2003} and eigenmodes~\cite{Vanfretti2011} have been estimated and inter-area oscillations~\cite{Vanfretti2010}, as well as power oscillation damping~\cite{Uhlen2012}, have been observed. 
Nevertheless, stochastic elements, more commonly denoted as ambient noise~\cite{Pierre1997}, are not in the focus of most power system studies. 
Only some studies include noise, e.g. when investigating power generation~\cite{Papaefthymiou2006}. 
Some recent papers have provided a detailed stochastic analysis and modelling of the bulk frequency without addressing spatial aspects of the dynamics~\cite{Schaefer2018,Gorjao2020,Anvari2020}.
Meanwhile, other works exist, highlighting the importance of short-term volatility in renewables \cite{Schmietendorf2017} and even first spatio-temporal considerations in Continental Europe \cite{Haehne2019}.

Both spatio-temporal or general time series analysis require access to high-quality, spatially distributed frequency recordings with phasor measurement units (PMUs) or PMU-like devices. 
Unfortunately, most data sets are not shared openly~\cite{Pfenninger2017b}. 
A discontinued university-based PMU network in the Nordic area was briefly in place between 2012 and 2014, which provides some of the datasets that are examined in this work~\cite{Almas2014}.
More recent initiatives only cover spatial measurements in Continental Europe~\cite{Gorjao2020,Jumar2020}.

Within this article, we present, analyse, characterise, and explain the aforementioned data of the Nordic synchronous area and study grid synchronisation in a data-centred model-free approach.
We focus on the increment statistics of the frequency time series which carries essential information on the fluctuations and the synchronisation of the frequency.
We show that non-Gaussian increment statistics are ubiquitous and that the variance on the increments scales faster than Brownian-like motions. 
Next, we show that stochastic fluctuations exhibit spatial correlations between locations even at vanishing time differences and that there exist temporal correlation within the same incremental time series. 
Finally, we examine phase and amplitude synchronisation separately.
To this end, we firstly observe phase synchronisation propagating linearly through the network, which contrasts the second finding of an diffusive-like coupling in amplitude synchronisation. 

\section{Synchronisation phenomena and increment statistics}

In this article, we analyse power-grid frequency recordings in the Nordic Grid from the 9\textsuperscript{th} to the 11\textsuperscript{th} of September 2013, with a sampling time of $0.02$ seconds. Note the approximately 36 hours of data available here already allow the analysis of the fluctuations on the time scale of seconds. 
In addition to the spatially well-defined measurements from 2013, we also evaluate the same September days from 2020 measured at an unknown location in central Sweden, provided by the Swedish TSO, obtaining qualitatively similar results (not shown).

We illustrate the locations of the recording sites on a map of the Nordic Grid synchronous area in Fig.~\ref{fig:trajectories}\textbf{a}.
The recordings were taken synchronously at six universities in Sweden and Finland, here abbreviated to acronyms: 
Chalmers University of Technology Gothenburg (CTH); Faculty of Engineering, Lund University (LTH); Royal Institute of Technology Stockholm (KTH); 
Luleå University of Technology (LTU); Tampere University of Technology (TTY); Aalto University (AU)~\cite{Almas2014}.
An excerpt of $30$ minutes of recordings is displayed in \textbf{b}, vertically displaced for clarity. 
Panel \textbf{c} shows a snapshot of $2$ seconds length, around the time point indicated by an arrow in panel \textbf{b}.
The fluctuations in TTY and AU show a smaller standard deviation than the remaining sites due to the larger number of generators in the Northern part of the synchronous area, compared to the mostly consumer-dominated south of the grid, see also Refs.~\cite{Schaefer2018,Gorjao2020b}.

Inspecting the time series, one immediately notices the most common properties of power-grid frequency: On course scales, all recordings are synchronous and seem perfectly identical at first sight.
If one examines them at a time scale of hundreds of milliseconds (in Fig.~\ref{fig:trajectories}\textbf{c}) the synchronous behaviour is still present, but small fluctuations occur, each seemingly with their own dynamics.
We will focus on these small fluctuations on time scales of a few seconds and introduce here the rather intuitive idea of phase and amplitude synchronisation.

\subsection{Phase and amplitude synchronisation}

\begin{figure}[t]
    \includegraphics[width=\linewidth]{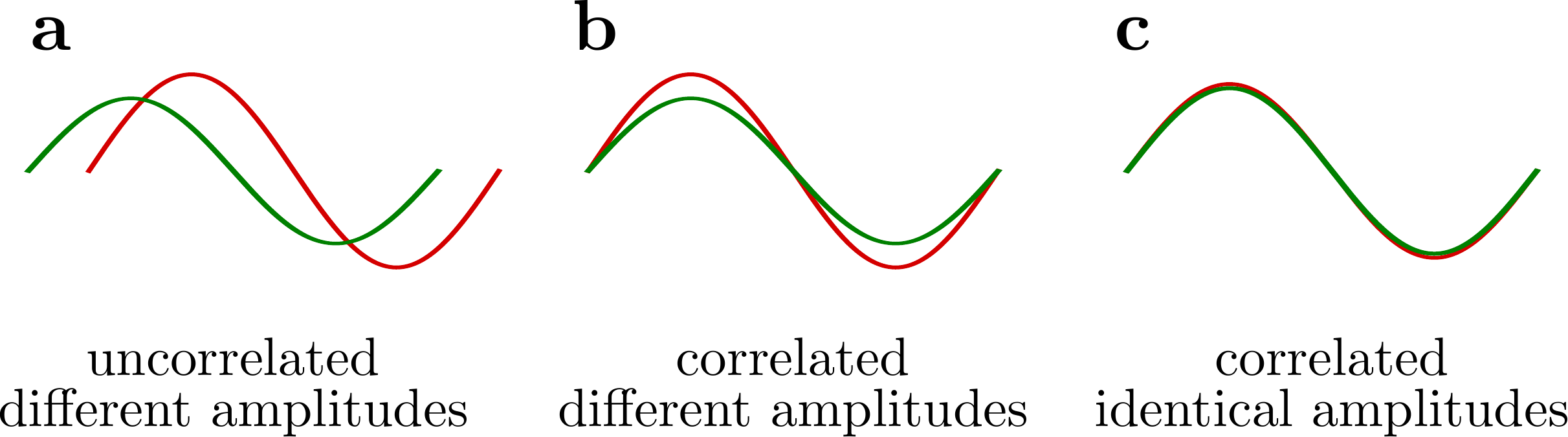}
    \caption{Synchronisation phenomena.
    Distinct frequency signals first become correlated and then converge to identical amplitudes. 
    We illustrate this here:
    \textbf{a}: Initially, signals are initially uncorrelated and have differing amplitudes.
    \textbf{b}: After some time, signals which are phase synchronised but still differ in amplitude.
    \textbf{c}: Finally, signals are synchronised both in phase and amplitude.}\label{fig:Synchronization_Illustration}
\end{figure}

A coupled dynamical system, say for simplicity, a two-area system, with one area exporting and the other area importing power, can be represented by a simplified model of two generators, 
each with given inertia, exchanging power, which displays some frequency dynamics around an equilibrium~\cite{Chompoobutrgool2013}.
Three different regimes of the collective dynamics exist, while the generators synchronise (typically towards $50/60\,$Hz), as depicted in Fig.~\ref{fig:Synchronization_Illustration} for two sinusoidal signals (e.g. a phase voltage at two different nodes of the power grid).
If both signals are only weakly interacting and independent, we would expect these signals to be uncorrelated, i.e., with a different phase and displaying different amplitudes (panel \textbf{a}). 
Enforcing synchronised dynamics, e.g. via coupling, will match their phases but not yet their amplitude.  
Hence, we observe a correlation between both signals (panel \textbf{b}). Lastly, given sufficient coupling, we also obtain amplitude synchronisation, i.e., in this case, a full synchronisation, as both phase and amplitude are equal (panel \textbf{c}).
Note that the amplitude discussed here is not the voltage amplitude but the amplitude of frequency fluctuations.

Naturally, several questions arise from this depiction: What are the usual timescales for each synchronisation to take effect after an external perturbation? 
What are the causes of this synchronisation?
How do distances between elements impact synchronisation (especially in networked systems)?
Lastly, are there properties from one synchronisation phenomenon that affect the other?

We will address these questions based on measured time series of power-grid frequency.
Phase synchronisation is analysed in terms of the signals' increments, i.e., the change of the signal between two points in time. 
If two signals are perfectly phase synchronised, their increments will be perfectly correlated. 
In contrast, amplitude synchronisation must be studied in terms of the original time series.

\subsection{Increment statistics}

To study the synchronisation, including correlations of each time series, we investigate the increment statistics of the recordings.
Synchronous power-grid systems operate at a set frequency.
In the Nordic Grid, and all of Europe, this is the nominal frequency of $50~\!$Hz. 
Many of the deterministic properties are studied directly from the time series themselves, by either studying their deterministic properties, i.e., as a dynamical system~\cite{Kundur2004},
or as a stochastic process~\cite{Milano2013,Jonsdottir2020}.
Yet, here we take another approach and focus on fluctuations of the time series---in particular their increments---and quantify their stochastic properties and correlations.

Increments $\Delta f_\tau(t)$ are defined as
\begin{equation}
\Delta f_\tau(t) =  f(t+\tau) - f(t),
\end{equation}
where the incremental lag $\tau$ gives the temporal difference of the two points within a time series $f(t)$.
Studying the incremental properties of a time series focuses on the shortest time scales of the underlying processes, thus excludes the deterministic trends and deals solely with the stochastic characteristics of the fluctuations themselves.
Note that we move away from considering the recordings in their time domain and study the difference between two points separated by a temporal lag $\tau$.

There is an intrinsic relation between increments and inertia in the system.
The ability of a power grid to maintain itself at its nominal frequency ($50\,$Hz or $60\,$Hz) is largely determined by the energy that can be stored and exchanged by the masses of the coupled rotating generators.
Short term fluctuations in the power system---i.e., precisely the increments---are conjectured to grow in amplitude and frequency when more inertia is removed as renewable generators replace fossil-fuelled ones~\cite{Ulbig2014,Milano2018}.
These fluctuations are of great importance as they can lead to large angle variations and power flow fluctuations, putting additional strain on generators and transmission system components.

After obtaining the incremental time series for all six sites, we investigate their statistics for the shortest incremental lag $\tau=0.02\,$s and the longer lag $\tau=1\,$s in Fig.~\ref{fig:trajectories}\textbf{d}-\textbf{f}. 
One observes considerable differences in the empirical probability distributions (PDFs) between the six sites. As a base-line, we might expect Gaussian distributions, i.e., inverted parabola in the vertical logarithmic plot (dotted line).
Indeed, inspecting the incremental distributions at a delay $\tau=1\,$s, some sites, such as CTH, approximately follow such a Gaussian distribution. 
In contrast, most sites display clear deviations from the expected Gaussian behavior for the short delay $\tau=0.02\,$s, i.e. they display heavy-tailed distributions.

The heavy tails in these distributions indicate that uncommonly large fluctuations take place in the increments, i.e., the difference from one time-point to the next is abnormally large (compared to normally distributed noise).
Furthermore, we note that these increments seem rather distinct at each location, in particular, compared to the much more homogeneous power-grid frequency recordings.
This tells us straightforwardly that, although there is strong synchrony in the system---the $50\,$Hz of operation---each location varies ever-so-slightly and each in its own manner.
What we observe here are the local properties of the recordings.
An in-depth analysis of the underlying mechanism giving rise to these distribution of the increments and their local properties is addressed in Ref.~\cite{Gorjao2021}.

To best quantify both the deviations from Gaussian distributions as well as the differences between sites, we investigate two statistical parameters: the variance and the kurtosis of each distribution, as a function of time step $\tau$.

\subsection{Statistical properties of incremental time series}

\begin{figure}[t]
    \includegraphics[width=\linewidth]{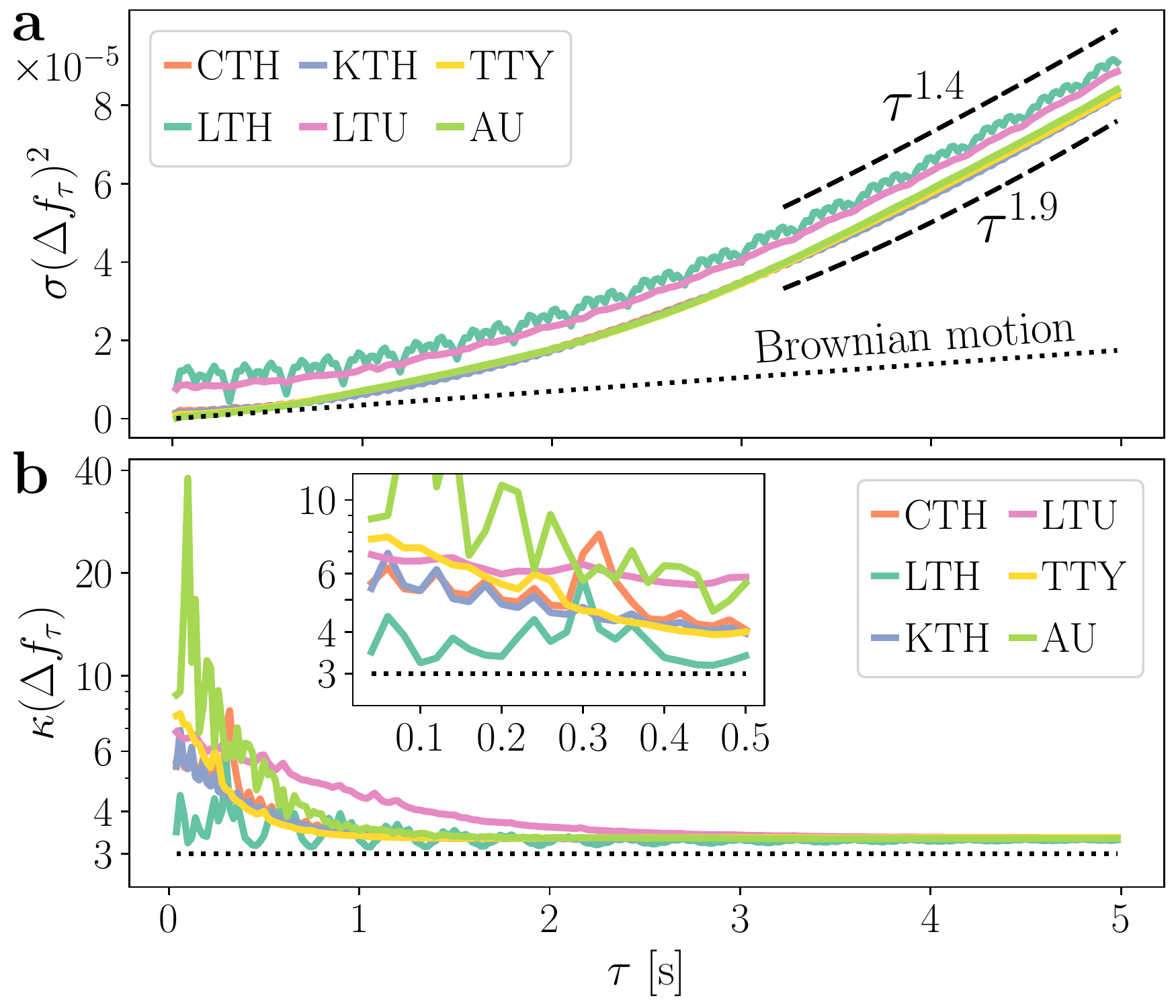}
    \caption{
    Variance $\sigma(\Delta f_\tau)^2$ and kurtosis $\kappa(\Delta f_\tau)$ of the increment $\Delta f_\tau$ display clear deviations from Gaussianity. 
    \textbf{a} A power-law scaling is observed, with exponent $>1$. For comparison, the linear-like scaling of an uncorrelated Brownian motion is shown. LTU and LTH seem to display the presence of microscopic noise, uniformly increasing the variance of their increments, irrespective of $\tau$.
    \textbf{b} Kurtosis $\kappa(\Delta f_\tau)$ of the increment $\Delta f_\tau$ in a vertical logarithmic scale. 
    The increments statistics are always leptokurtic, i.e., $\kappa>3$. All increment statistics $\kappa(\Delta f_\tau)$ converge to $\kappa(\Delta f_{\tau\gg 0}) \approx 3.35$ at $\tau\gtrsim2$.}
    \label{fig:var_kurtosis}
\end{figure}

We examine two statistical moments of the incremental distributions as a function of the incremental lag $\tau$, namely, the variance and kurtosis.
The variance (i.e., the second moment with subtracted mean) indicates the average displacement of each recording from their mean, i.e., how far the point of the recordings are spread out from their mean. Note that we have a mean zero here, as we investigate frequency increments.
On the other hand, the kurtosis (fourth normalised moment) roughly indicates how often rather large deviations happen, i.e., deviations much larger than those that fall inside the spread of the variance~\cite{Westfall2014}.

Note that we handed-picked two distributions at $\tau=0.02\,$s and $\tau=1\,$s for Fig.~\ref{fig:trajectories}\textbf{d}-\textbf{f}.
Now, we systematically investigate how the variance and kurtosis of each increment distribution change as we increase the incremental lag $\tau$.
Let us examine the variance $\sigma(\Delta f_\tau)^2$ (Fig.~\ref{fig:var_kurtosis}\textbf{a}) and the kurtosis $\kappa(\Delta f_\tau)$ (Fig.~\ref{fig:var_kurtosis}\textbf{b}) for the first five seconds of incremental lags $\tau\in[0.02\,$s$,5\,$s$]$.
The first observation is that the variance of the increments $\Delta f_\tau$ increases in a power-like relation.
In particular, we observe a scaling of the variance as
\begin{equation}\label{eq:var_scaling}
    \sigma(\Delta f_\tau)^2 \sim \tau^{2H}, ~\mathrm{with}~ 1.4<2H<1.9,
\end{equation}
which we can quantify using the Hurst exponent $H$~\cite{Hurst1951}. We know that the increments of a time series are intrinsically related to the nature of the fluctuations, i.e., the ambient noise. For example, classical \textit{uncorrelated} Brownian motion with mean zero---often assumed when simulating noisy processes, e.g. as in Ref.~\cite{Pierre1997}---scales with $\sigma(\Delta_{\mathrm{Bm}} f_\tau)^2 \sim \tau$, i.e., $H=0.5$, which is given by the dotted line in Fig.~\ref{fig:var_kurtosis}\textbf{a}. 

What we observe here is a strong diffusion scaling of the fluctuations. It is  particularly relevant to take this scaling into account when performing any from of simulations of the power-grid frequency and thus should be incorporated accordingly \cite{Jonsdottir2020}. 
Importantly, the observation of this correlated and non-white noise is ubiquitous and uniform across locations and evidenced from a data analysis, i.e., without imposing any model.
We come back to this observation in Sec.~\ref{sec:amp_synch}, precise the value of the Hurst coefficient $H$.

As a last remark, one observes similarly that in both LTU and LTH additional microscopic noise is present, possibly with some regular properties (notice the curves do not approach zero for decreasing $\tau$).
Whether these are artefacts or fundamental physical property of each site's local stochastic properties is left to future analyses.

Let us now turn our attention to how the kurtosis changes with increasing $\tau$ (Fig.~\ref{fig:var_kurtosis}\textbf{b}). As a reference, we provide the kurtosis of a normal distribution as $\kappa_\mathcal{N}=3$ in the plot. 
Notably, the kurtosis of the increments decreases with growing incremental lags $\tau$, that is to say, as $\tau$ increases the distributions become more and more like normal distributions. 
One should note that this is not completely unexpected behaviour. 
Recent observations by Hähne \textit{et al.} \cite{Haehne2019} and Schmietendorf \textit{et al.} \cite{Schmietendorf2017} point towards an increase of this phenomena generated by volatile wind-energy input.
This also fundamentally links properties of power-grid frequency with, e.g turbulent flows~\cite{Castaing1990,Castaing1994,Beck2003} and stock market prices~\cite{Xu2016}.
In particular, nontrivial scaling of the probability densities depending on the time lag parameter $\tau$ and involving a fractional Brownian motion parameter $H$ was experimentally found in turbulent Taylor--Couette flow~\cite{Beck2001b}. 
In that sense, the frequency difference fluctuations on a small scale have similarities with those of small-scale velocity differences in turbulent flows, at least in a statistical sense.
What is relevant for our analysis is to understand this as a synchronisation phenomenon.
As depicted in Fig.~\ref{fig:Synchronization_Illustration}\textbf{c}, if one is to achieve amplitude synchronisation, then obviously two recordings must end up with identical distributions.
This does not mean that the kurtosis of all incremental time series must become normally distributed but does imply they must be identical.

With these insights of the increment statistics at each location, we can now address the first guiding question formulated in the introduction: Can we unravel the physics of phase synchronisation and its timescale? To do so we must study the smallest timescales of the incremental time series and focus on their correlations.

\subsection{Phase synchronisation}

\begin{figure}[t]
    \includegraphics[width=\linewidth]{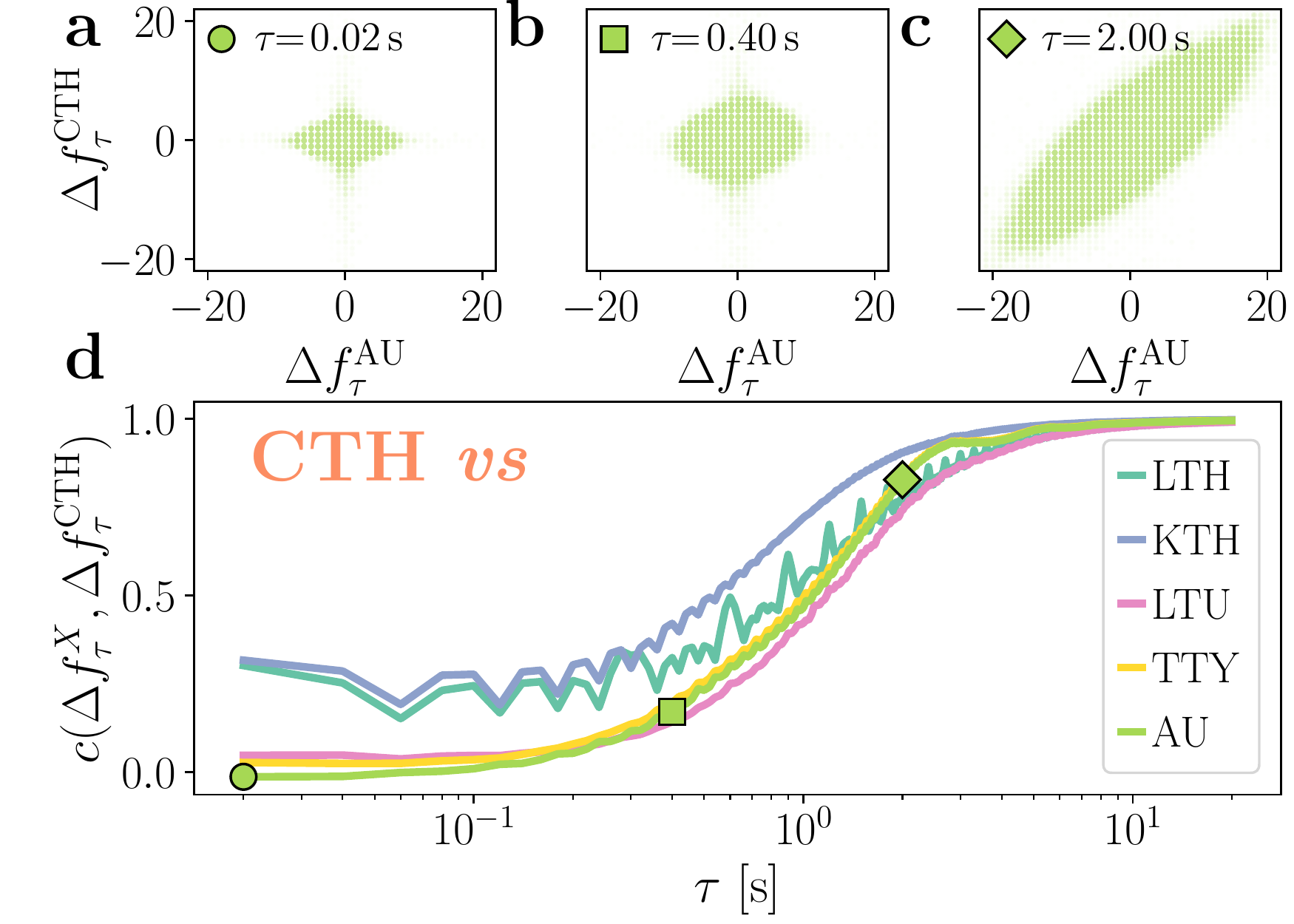}
    \caption{Correlations of the increments increases with delay $\tau$.
    (\textbf{a}-\textbf{c}) display the correlations between CTH and AU at \textbf{a} $\tau=0.02\,$s, \textbf{b} $\tau=0.40\,$s, and \textbf{c} $\tau=2.00\,$s.
    (\textbf{d}) Displays the Pearson correlation $c(\Delta f_\tau^X,\Delta f_\tau^{\mathrm{CTH}})$ in $\tau\in[0.02\,$s$,20\,$s$]$.
    Generally correlations are zero or small with all other recordings at $\tau=0.02s$, yet do not vanish for the closest locations to CTH: LTH and KTH. For larger increments, the correlation approaches one: $\lim_{\tau \rightarrow \infty}c=1$.}\label{fig:Pearson_CTH}
\end{figure}

Let us examine the emergence of phase synchronisation of the incremental time series (from uncorrelated to correlated time series).
Power grids are designed to ensure that all power generators work synchronously across the power grid.
Naturally---and as we have seen so far---this means that the frequency recordings themselves are almost identical, i.e., highly correlated with each other, yet this does not directly translate into fluctuations at each location acting similarly.
Indeed, both models and theory treating the power-grid frequency often assume uncorrelated fluctuations.
Again without relying on any model, we will show that spatial correlations of the fluctuations are ubiquitous, and especially prevalent below a certain distance---similar to a scaling law in long-range fluctuations in power-grid frequency presented in Ref.~\cite{Gorjao2020b}.

To study the correlations of the increments between two locations, we examine the Pearson correlation of two recordings $c(\Delta f_\tau^X,\Delta f_\tau^Y)$
\begin{equation}\label{eq:pearson}
    c(\Delta f_\tau^X,\Delta f_\tau^Y) =\frac{ \mathrm{Cov}\!\left(\Delta f_\tau^X,\Delta f_\tau^Y\right)}{\sigma(\Delta f_\tau^X)\sigma(\Delta f_\tau^Y)} \in [-1,1],
\end{equation}
with $X$ and $Y$ the two locations or recordings, $\sigma(\cdot)$ their individual standard deviation and $\mathrm{Cov}(\cdot,\cdot)$ their covariance.
A value of $c=1$ indicates total correlation, $c=-1$ total anti-correlation, and $c=0$ no correlation.

Let us take the site CTH as an example.
The correlation $c(\Delta f_\tau^X,\Delta f_\tau^{\mathrm{CTH}})$ with all other locations is displayed in Fig.~\ref{fig:Pearson_CTH}, for $0.02\,$s $<\tau<20\,$s. Panels \textbf{a-c} show how large incremental lags $\tau$ lead to highly aligned, i.e., correlated increments, while short incremental lags $\tau$ display no correlation. 
The de-correlation of the increments for very small time differences $\tau$ is clear, yet this does not seem to be the case for the two closest locations to CTH: LTH and KTH.
In this case, the Pearson correlations are $c(\Delta f_\tau^\mathrm{LTH},\Delta f_\tau^\mathrm{CTH})\sim 0.25$ and $c(\Delta f_\tau^\mathrm{KTH},\Delta f_\tau^\mathrm{CTH})\sim 0.25$ at the lower limit of $\tau\to 0$.

\begin{figure}[t]
    \includegraphics[width=\linewidth]{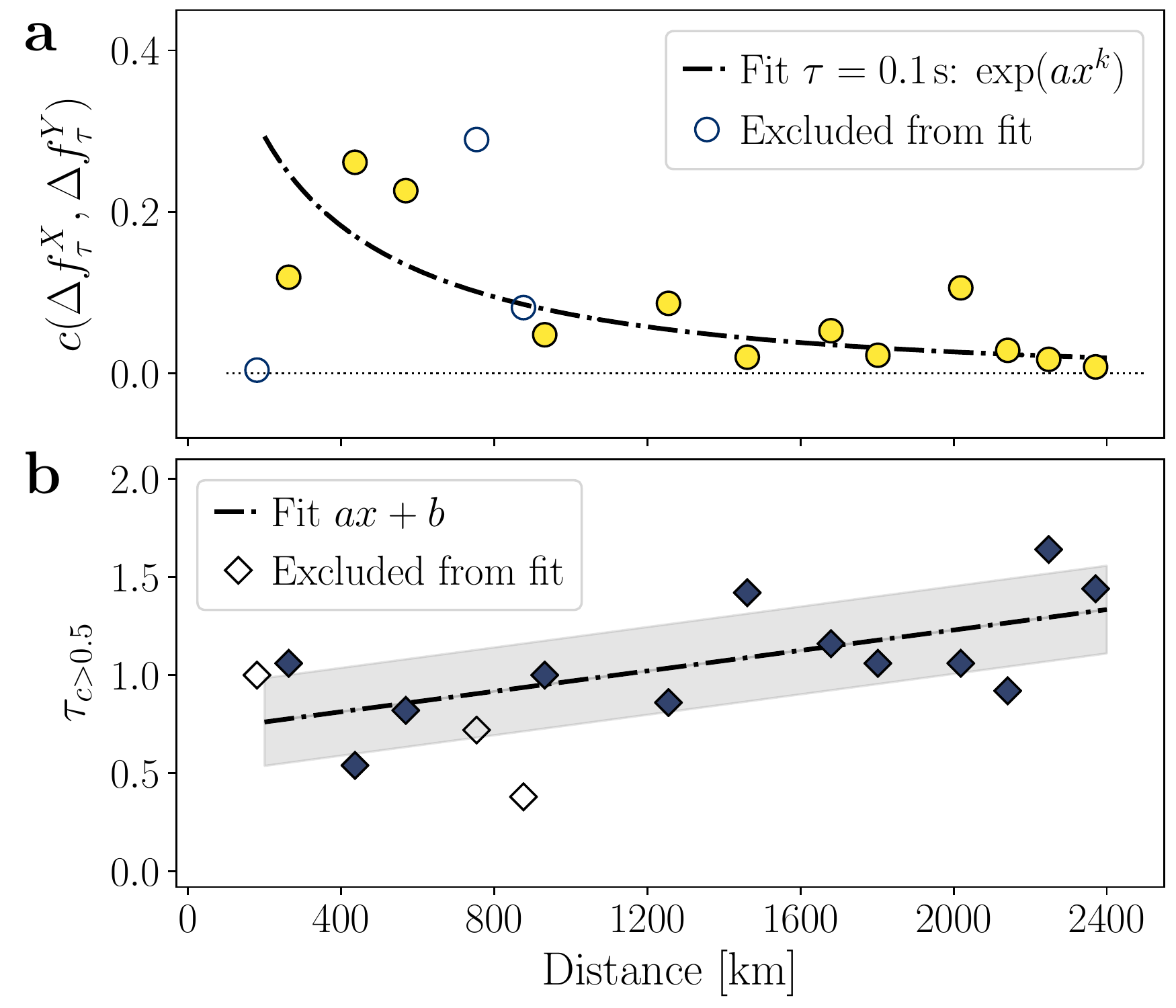}
    \caption{
    Time to reach a correlated state increases approximately linearly with distance.  
    (\textbf{a}) displays the Pearson correlation $c(\Delta f_\tau^X,\Delta f_\tau^Y)$ of two locations at the temporal lag $\tau = 0.1\,$s in relation to their driving distance.
    The presence of non-vanishing correlations at small distances can be observed, following an exponential-like function exp($ax^k$), with parameters $a=-0.10$ and $k=0.47$.
    (\textbf{b}) displays the time at which the Pearson correlations between two locations become greater than $0.5$, $\tau_{c>0.5}$, plotted against the distance between the locations.
    In similar fashion, we see that the time it takes for the increments to become correlated is approximately linearly proportional to the distance between two locations.
    Shaded area indicates the standard deviation of the polynomial fitting.
    Three couplings are discarded: AU--TTY and AU--LTU, due to their seemingly small anti-correlation at $\tau = 0.02\,$, which quickly becomes correlated at $\tau>0.1s$, and the LTH--KTH coupling, which as seen in Fig.~\ref{fig:var_kurtosis}\textbf{a} presents microscopic noise, mooting possible correlations.}\label{fig:Pearson_distance}
\end{figure}

This gives rise to the following question: Are fluctuations in the increments correlated for shorter distances, i.e., are there spatial correlations in the increments themselves?
We examine this by studying the Pearson correlations against the geographical distance between the locations in two manners: (1) find the correlation at the lowest temporal lag $\tau=0.1\,$s and the distance between the locations; (2) find how much time it takes for the locations to surpass a given threshold of correlation and the distance between the locations.

Firstly we note that all sea cables in the Baltic Sea or the Gulf of Bothnia are direct current (DC) cables, they transmit power asynchronously.  
That being the case, driving/walking distances between two locations serve as a proxy to actual electric distance, i.e., the equivalent AC impedance between two locations~\cite{Cotilla2012}.

In Fig.~\ref{fig:Pearson_distance}\textbf{a}, we display the fifteen distance pairs between the six locations and their respective Pearson correlation at temporal lag $\tau = 0.1\,$s are displayed.
The Pearson correlations $c(\Delta f_\tau^X,\Delta f_\tau^Y)$ for all locations at $\tau=0.1\,$s are plotted against the distances between the points.
A pattern of correlations emerges for short distances with non-vanishing Pearson correlations.
Three couplings are discarded: (1) AU--TTY and AU--LTU, due to their small anti-correlation at very small $\tau = 0.02\,$s--$0.1\,$s.
(2) the LTU--LTH coupling, where the present microscopic noise moots evaluating correlations (see e.g. Fig.~\ref{fig:var_kurtosis}\textbf{a}).

In Fig.~\ref{fig:Pearson_distance}\textbf{b} we examine the speed at which correlations between two locations become larger than $0.5$ ($50\%$), i.e., we find the time $\tau_{c>0.5}$ at which the correlations $c(\Delta f_\tau^X,\Delta f_\tau^Y)>0.5$ in relation to the distance between the locations.
We observe that locations closer to each other see their increments becoming correlated faster than locations that are further apart.
While this general trend is expected, note the approximately linear relation between phase-synchronisation and geographical distance between the sites.
We will have cause to contrast this with amplitude synchronisation in the following section.
Furthermore, notice that this phenomenon is strictly bound to be taking place at a time scale $<2\,$s, i.e., phase synchronisation takes place at very short temporal scales.
These results point towards a potential linear phase-synchronisation phenomenon in power-grid frequency recordings, which should be carefully examined with more spatially-extended data.

We now move forward to examine the emergence of amplitude synchronisation.

\subsection{Amplitude synchronisation}\label{sec:amp_synch}

We have seen that phases of increments synchronised within $<2\,$s, so let us now move to amplitude synchronisation.
As we have seen in Fig.~\ref{fig:var_kurtosis}\textbf{a} the variance of the increment statistics increases in a power-law relation with the incremental lag $\tau$.
In order to best describe this---and to retrieve the Hurst exponent $H$---we employ Detrended Fluctuation Analysis (DFA) of the power-grid frequency recordings~\cite{Peng1994,Kantelhardt2002,Gorjao2021b}.
Recall that DFA studies the scaling of the fluctuations of a time series by studying the local properties of the data, similarly to what increment statistics does.
With DFA we extract the fluctuation function $F(r)$ over a scale $r$ of a time series, in a much similar fashion to our incremental lag $\tau$.
We will keep $\tau$ and $r$ distinct since the DFA is applied directly to the power-grid frequency recordings, not the incremental time series.
Still, DFA and increment analysis are intrinsically related and studying the fluctuation function $F(r)$ as a function of the scale $r$ will help us uncover the scaling parameter $\alpha$, as also observed in Fig.~\ref{fig:var_kurtosis}\textbf{a}.

The DFA procedure, applied to the power-grid frequency recordings, is the following: Take non-overlapping segments of the power-grid frequency, fit a polynomial function (of order one, in this case), subtract the fit from the segment of data, extract the variance and average over each detrended segment.
By increasing the segment size one can study the change of the variances, i.e., fluctuation function $F(r)$, as a function of the segment size (the scale $r$).
If the time series follows a power-law, which we show in Fig.~\ref{fig:var_kurtosis}\textbf{a} it does, one can evaluate the plots of the fluctuation function $F(r)$ in the scale $r$ in a double logarithmic scale.
The reason to compute the fluctuation function $F(r)$ instead of examining the original time series is simple: 
Actual power-grid frequency recordings have trends, such as short-terms jumps due to dispatch and market activity or a permanent small mismatch of power generation and consumption.
We wish to disentangle these trends from the true underlying fluctuation dynamics, which DFA is capable of.

\begin{figure}[t]
    \includegraphics[width=\linewidth]{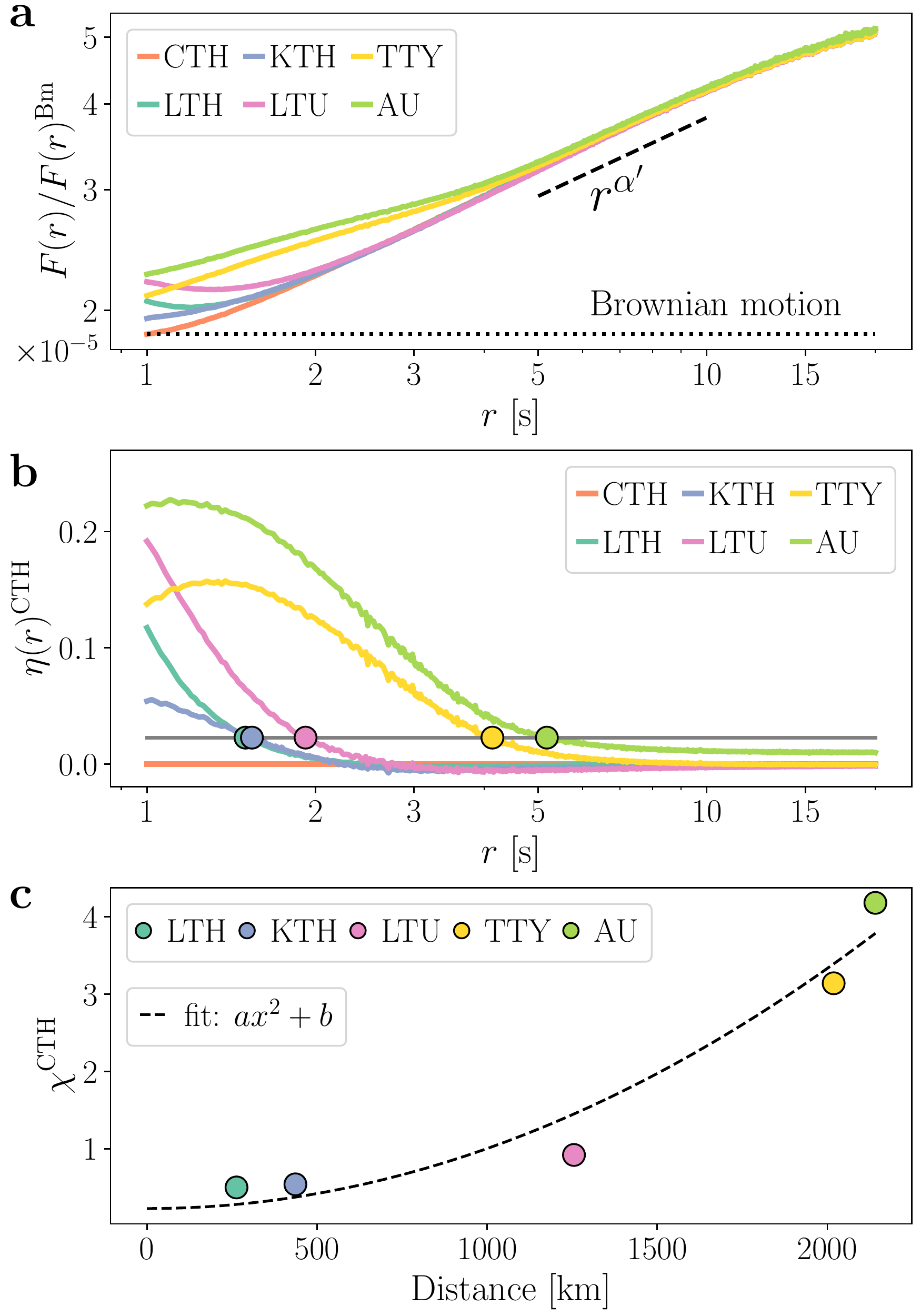}
    \caption{
    Correlated, diffusive coupling revealed by DFA, relative DFA $\eta(r)$, and `time-to-bulk' $\chi$.
    (\textbf{a}) displays the normalised fluctuation function $F(r)$ obtained from DFA.
    normalised by the fluctuation function $F(r)^{\mathrm{Bm}}$ of a Brownian motion ($\alpha_{\mathrm{Bm}}-1=0.5$).
    A fit to the empiric fluctuation function yields the scaling exponents $\alpha^\prime = \alpha - \alpha_{\mathrm{Bm}} = 0.376$ or equivalently a Hurst index $H=\alpha-1=0.876$, i.e., strongly positively correlated noise.
    (\textbf{b}) displays the relative fluctuation function $\eta(r)^{\mathrm{CTH}}$, with CTH as reference.
    The relative fluctuations of the other five locations decay in time.
    The markers indicate the points (in time) where the relative size of the fluctuations hits 0.1 (10\%) of the maximal relative fluctuation.
    (\textbf{c}) shows the `time-to-bulk' function $\chi^{\mathrm{CTH}}$ against the distances of each recordings' location in relation to CTH.
    A quadratic fit (thus diffusive coupling with distance) approximates the amplitude synchronisation phenomenon.}
    \label{fig:DFA}
\end{figure}

For our purpose here, we have already established that the incremental time series follows a diffusion scale with a Hurst exponent $H>0.5$ given by Eq.~\eqref{eq:var_scaling}, which is manifestly larger in exponent than an uncorrelated Brownian motion.
We plot the fluctuation function $F(r)$ of the six time series in a scale $r\in[1\,$s$,20\,$s$]$ in Fig.~\ref{fig:DFA}\textbf{a}, normalised by the scaling of a Brownian motion, which has a scaling power $\alpha_{\mathrm{Bm}}=1.5$ (i.e., $\alpha = H+1$).
For illustrative purposes, we show the difference in scaling, represented by $\alpha^\prime = \alpha - \alpha^{\mathrm{Bm}}$.
We find that $\alpha^\prime\approx0.376$, i.e., $\alpha = 1.876$, for the range of $r\in[5\,$s$,10\,$s$]$.
Noticeable is also the distinct separation of the curves at timescales $r<5\,$s, which has been identified in Ref.~\cite{Gorjao2020b} as a possible method to study the synchronisation of fluctuation over spatial distances.
In order to compare the fluctuation functions $F(r)$ an each location, i.e., to fundamentally compare the amplitudes of the fluctuations at each locations, we compute the relative DFA function $\eta(r)$, as introduced in Ref.~\cite{Gorjao2020b}, with CTH as a reference in Fig.~\ref{fig:DFA}\textbf{b}.
This function represents the relative variations of the fluctuations to a reference point, given by
\begin{equation}
    \eta(r)^Y = \frac{F(r)^X - F(r)^Y}{F(r)^Y},
\end{equation}
with $Y$ the reference location, CTH in our case, and $X$ the remaining locations.
The synchronous nature of power-grids ensures all fluctuations eventually collapse into a single function.
That means all frequency increment amplitudes become identical.

To quantify this synchronisation, we define a `time-to-bulk' $\chi$ as follows:
We record the time it takes each time series' relative fluctuations $\eta$ to be reduced to $0.1$ ($10\%$) and thereby become indiscernible from the reference (indicated by the grey line in Fig~\ref{fig:DFA}\textbf{b}). 
Using CTH as a reference, we obtain $\chi^{\mathrm{CTH}}$ and observe how the time towards synchronisation increases with distance in Fig.~\ref{fig:DFA}\textbf{c}.
What we observe here contrasts sharply with the linear relation of phase synchronisation in Fig.~\ref{fig:Pearson_distance}.
Here we observe an approximately diffusive coupling, i.e., the time it takes for the amplitude of the increments to become identical grows approximately with the square of the distance, as described in Ref.~\cite{Gorjao2020b}. 
Interestingly, the data suggest a potentially even stronger dependency with distance, which could be linked to strong temporal correlations discussed earlier: If the dynamic at one site depends substantially on its own past (high temporal correlation), then it will take longer to synchronise fully with the remaining network. 
This poses many interesting research questions for the future, emphasising the importance of correlations.

We pause here to thoroughly examine what the `time-to-bulk' analysis just uncovered.
First, we confirmed the suspicion raised in Fig.~\ref{fig:var_kurtosis}\textbf{a} that the power-grid frequency indeed exhibits temporally correlated noise, i.e., the assumption that the high-frequency fluctuations in power-grid frequency recordings are purely white noise is not justified.
In fact, $\alpha>1.5~(H>0.5)$ indicates the presence of positively correlated motions, which show a power-law like diffusion, as seen in Fig.~\ref{fig:var_kurtosis}\textbf{a}.
One finds that all time series coalesce to a Hurst index of $H=\alpha-1=0.876$, i.e., strongly positively correlated noise.
This Hurst index can be determined efficiently for any given power-grid frequency measurement and should be incorporated in stochastic studies of power systems. 

Secondly, we observe that these fluctuations display a spatial relation between the locations, i.e., locations that are closer see their fluctuation amplitude synchronise faster than those farther apart. 
Note that this is a second effect, different from the (temporal) correlation in the time series, which we have examined before.
Focusing on the new spatial correlations, we know all power-grid frequency recordings are highly correlated, they practically follow the same phenomena at timescales $>5\,$s, i.e., their dynamical oscillations around $50\,$Hz.
Moreover, we have seen above that the correlations of the increments are already large at these long timescales, while displaying zero or low correlations on short time scales.
What we now observe is a synchronisation of the amplitudes of the fluctuations as a function of time.
If we compare Fig.~\ref{fig:DFA} to Fig.~\ref{fig:Pearson_distance}, we notice a sharp difference in the physics of the synchronisation phenomena: The amplitude synchronisation (see Fig.~\ref{fig:DFA}) is achieved approximately via diffusive coupling, i.e., following a quadratic relation with distance, in sharp contrast to the linear relation for phase synchronisation (see Fig.~\ref{fig:Pearson_distance}).

We note that \textit{a priori} there is no necessary relation between local temporal properties of the incremental time series and the spatial convergence of all oscillations into one bulk behaviour.
Nevertheless, the fact that local temporal properties affect the spatial correlations of systems that rely heavily on synchronisation to operate is plausible.
This, however, has not been described before.

\section{Conclusion}
In this article, we have analysed synchronous recordings of the  power-grid frequency from six locations in the Nordic Grid from September 2013.
The high temporal resolution of these recordings of $0.02\,$s allows a detailed analysis of power system synchronisation via the statistics and correlations of the increments.
Essential insights into the temporal and spatial scales of synchronisations can be extracted from the ambient fluctuations in a model-free approach.
Our results further emphasise the outstanding importance of broad availability of high-quality data for research on power system operation and energy science in general~\cite{Pfenninger2017}.

We investigated the distribution of increments ($\Delta f_\tau(t) = f(t+\tau)-f(t)$) and noted severe deviations from Gaussianity. In particular, increment distributions are highly leptokurtic, i.e., display heavy tails. 
Noticeable differences in the increment statistics at every location are observed, indicating that the incremental time series reflect above all the local phenomena of power generation and consumption in the location of the recording. 
We also saw a relaxation of the incremental time series kurtosis for large time lags $\tau$, yet even for very large  lags we observe a leptokurtic distribution with kurtosis of $\kappa(\Delta f_{\tau\gg0}) \approx3.35$. We continued with a detailed discussion of phase and amplitude synchronisation of frequency increments.

The emergence of phase synchronisation is revealed by the correlations of the increments on different temporal and spatial scales. 
Our analysis has shown two essential results:
(1) If locations are relatively close electrically, the increments are correlated even at the shortest possible time lag of $0.02\,$s, indicating that ambient fluctuations of power generation and consumption that drive the frequency dynamics are correlated.
In light of this result, any assumption of spatially independent noise in power system simulations should be carefully reviewed.
(2) The increments at locations become strongly correlated on a time scale of one second depending on the electrical distance of the location. 
The further two locations are from each other, the longer it takes before correlations exceed a certain value.

Strong temporal correlations in the fluctuations are revealed by the variance of the incremental time series.
We find that the variance follows a power-law $\sim\tau^{2H}$ as a function of the incremental lag $\tau$, with an exponent much higher than for ordinary uncorrelated Brownian motion. 
This result is confirmed by detrended fluctuation analysis (DFA), which yields the exponent $\alpha=1.876$, i.e., a positively correlated Hurst index $H=0.876$.
DFA is further used to study the time scales of amplitude synchronisation by quantifying the time needed until the fluctuation functions at two locations become similar up to a certain level.
This `time-to-bulk' function increases with distance approximately following diffusive coupling (quadratic function), as observed before in Ref.~ \cite{Gorjao2020b}. 
Critically, we note a systematic deviation from pure diffusive coupling, which suggests an even stronger scaling with distance, i.e. far-apart sites take longer to fully synchronise.
This finding is coherent with the observation of temporally correlated noise at each site: local temporal properties might influence the global synchronisation phenomena, delaying a fully synchronised power grid~\cite{Kettemann2016}.
Interestingly, the linear scaling of phase synchronisation, compared to the approximately quadratic scaling of amplitude synchronisation implies that locations far apart will first synchronise in their phase, then amplitude, while the ordering might be reversed for locations geographically nearby.
A definite answer will require data from more locations, preferably in large synchronous grids.  

We conclude that the implications of these findings are considerable for both the understanding of the dynamical processes in power grids as well as their simulation, as the increments provide a proxy for fluctuations of the power imbalance that drives the frequency dynamics.
Foremost, we uncovered that the temporal stochastic properties at each recording site impact the speed at which amplitude synchronisation is achieved.
This grants an exact measure for amplitude synchronisation across any power-grid, which can now be uncovered solely from one single local recording, i.e., since the scaling parameter $\alpha$ is a local property, a single power-grid frequency measurement in a synchronous region allows us to determine at which speed an electrically distant far-away location will be amplitude-synchronised with the rest of the grid.
Furthermore, two distinct and so far practically unaddressed characteristics are uncovered: First, fluctuations at different locations are correlated.
This implies immediately that any simulation---especially for small power grids as microgrids---must consider the presence not only of noise, i.e., stochastic fluctuations but of spatially correlated noise.
We observe that substantial correlations are seen up to distances of $1000\,$km.
Secondly, each location, each power-grid frequency recordings, shows a distinct strong temporal correlation within itself.
This is a clear indication that temporal correlations in stochastic fluctuations are present.
In fact, these correlations determine the physics of the aforementioned amplitude synchronisation.
Thus, adequate simulations should extent their analysis far beyond classical white noise (uncorrelated Brownian noise) and consider instead spatio-temporally correlated (non-white) noise.

All of these effects significantly impact risk assessment: heavy-tailed noise leads to larger deviations than expected from Gaussian noise, while correlated noise continues to push the system in one given direction, instead of randomly fluctuating around zero.
Hence, both correlations and non-Gaussian distributions induce larger frequency deviations than assumed from white Gaussian noise and thereby increase the risk of destabilising the system.
Simply adding a larger security margin when carrying out simulations is barely appropriate: a broad Gaussian noise distribution would imply that medium deviations occur very often when instead rare large deviations are the problem.
Hence, to properly dimension back-up capacities and design adequate control options, the non-standard statistics presented here should be taken into account.

In the future, it would be desirable to study the link between local temporal correlation and spatial amplitude synchronisation in different data sets~\cite{Klockner2017}.
Also, further theoretical work is necessary to offer efficient simulation and modelling tools for all discussed phenomena. 

\begin{acknowledgments}
We gratefully acknowledge support from the German Federal Ministry of Education and Research (grant no.~03EK3055B) and the Helmholtz Association (via the joint initiative ``Energy System 2050 -- A Contribution of the Research Field Energy'' and the grant ``Uncertainty Quantification -- From Data to Reliable Knowledge (UQ)'' with grant no.~ZT-I-0029).
This work was performed as part of the Helmholtz School for Data Science in Life, Earth and Energy (HDS-LEE).
This project has received funding from the European Union's Horizon 2020 research and innovation programme under the Marie Skłodowska-Curie grant agreement No 840825.
Luigi Vanfretti was supported in part by the Engineering Research Center Program of the National Science Foundation and the Department of Energy under Award EEC-1041877 and by the CURENT Industry Partnership Program.

\end{acknowledgments}

\bibstyle{apsrev4-2}
\bibliography{bib}

\end{document}